\documentstyle[epsfig]{mn}

\newif\ifAMStwofonts
                                                                                                                                                     
\ifoldfss
  \ifCUPmtlplainloaded \else
    \NewTextAlphabet{textbfit} {cmbxti10} {}
    \NewTextAlphabet{textbfss} {cmssbx10} {}
    \NewMathAlphabet{mathbfit} {cmbxti10} {} 
    \NewMathAlphabet{mathbfss} {cmssbx10} {} 
  \fi
  \ifAMStwofonts
    \ifCUPmtlplainloaded \else
      \NewSymbolFont{upmath} {eurm10}
      \NewSymbolFont{AMSa} {msam10}
      \NewMathSymbol{\upi}     {0}{upmath}{19}
      \NewMathSymbol{\umu}     {0}{upmath}{16}
      \NewMathSymbol{\upartial}{0}{upmath}{40}
      \NewMathSymbol{\leqslant}{3}{AMSa}{36}
      \NewMathSymbol{\geqslant}{3}{AMSa}{3E}

       \let\le=\leqslant
       
    \fi
  \fi
\fi 
\ifnfssone
  \newmathalphabet{\mathit}
  \addtoversion{normal}{\mathit}{cmr}{m}{it}
  \addtoversion{bold}{\mathit}{cmr}{bx}{it}
  \newmathalphabet{\mathbfit} 
  \addtoversion{normal}{\mathbfit}{cmr}{bx}{it}
  \addtoversion{bold}{\mathbfit}{cmr}{bx}{it}
  \newmathalphabet{\mathbfss} 
  \addtoversion{normal}{\mathbfss}{cmss}{bx}{n}
  \addtoversion{bold}{\mathbfss}{cmss}{bx}{n}
  \ifAMStwofonts
    \ifCUPmtlplainloaded \else
      \UseAMStwoboldmath
      \makeatletter
      \new@mathgroup\upmath@group
      \define@mathgroup\mv@normal\upmath@group{eur}{m}{n}
      \define@mathgroup\mv@bold\upmath@group{eur}{b}{n}
      \edef\UPM{\hexnumber\upmath@group}
      \new@mathgroup\amsa@group
      \define@mathgroup\mv@normal\amsa@group{msa}{m}{n}
      \define@mathgroup\mv@bold\amsa@group{msa}{m}{n}
      \edef\AMSa{\hexnumber\amsa@group}
      \makeatother
      \mathchardef\upi=''0\UPM19
      \mathchardef\umu=''0\UPM16
      \mathchardef\upartial=''0\UPM40
      \mathchardef\leqslant=''3\AMSa36
      \mathchardef\geqslant=''3\AMSa3E

       \let\le=\leqslant

    \fi
  \fi
\fi 
                                                                                                                                                     
\ifnfsstwo
  \DeclareMathAlphabet{\mathbfit}{OT1}{cmr}{bx}{it}
  \SetMathAlphabet\mathbfit{bold}{OT1}{cmr}{bx}{it}
  \DeclareMathAlphabet{\mathbfss}{OT1}{cmss}{bx}{n}
  \SetMathAlphabet\mathbfss{bold}{OT1}{cmss}{bx}{n}
  \ifAMStwofonts
    \ifCUPmtlplainloaded \else
      \DeclareSymbolFont{UPM}{U}{eur}{m}{n}
      \SetSymbolFont{UPM}{bold}{U}{eur}{b}{n}
      \DeclareSymbolFont{AMSa}{U}{msa}{m}{n}
      \DeclareMathSymbol{\upi}{0}{UPM}{``19}
      \DeclareMathSymbol{\umu}{0}{UPM}{``16}
      \DeclareMathSymbol{\upartial}{0}{UPM}{``40}
      \DeclareMathSymbol{\leqslant}{3}{AMSa}{``36}
      \DeclareMathSymbol{\geqslant}{3}{AMSa}{``3E}

       \let\le=\leqslant

    \fi
  \fi
\fi 
                                                                                                                                                     
\ifCUPmtlplainloaded \else
  \ifAMStwofonts \else 
    \def\upi{\pi}
    \def\umu{\mu}
    \def\upartial{\partial}
  \fi
\fi

\title{The CMB cold spot: texture, cluster or void?}

\author[M. Cruz et al.]{M. Cruz,$^1$\thanks{e-mail:
cruz@ifca.unican.es}
E. Mart\'{\i}nez-Gonz\'alez,$^1$
P. Vielva,$^1$ 
J. M. Diego,$^1$
M. Hobson,$^2$
N. Turok$^3$\\
$^1$IFCA, CSIC-Univ. de Cantabria, Avda. los Castros, s/n, 39005-Santander,Spain.\\
$^2$Astrophysics Group, Cavendish Laboratory, J.J. Thomson Avenue, Cambridge CB3 OHE, UK.\\
$^3$DAMTP, CMS, Wilberforce Road, Cambridge, CB3 0WA, UK.}

\date{Accepted  Received ; in original form }

\begin{document}

\maketitle

\begin{abstract}

The non-Gaussian cold spot found in the WMAP data has created controversy about its origin.
Here we calculate the Bayesian posterior probability ratios for three different models that could explain 
the cold spot. A recent work claimed that the Spot could be caused by a 
cosmic texture, while other papers suggest that it could be due to the gravitational effect produced by an 
anomalously large void. Also the Sunyaev-Zeldovich effect caused by a cluster is taken into account as a possible 
origin. We perform a template fitting on a $20 \degr$ radius patch centered at Galactic coordinates 
($b = -57^\circ, l = 209^\circ$) and calculate the posterior probability ratios for the void and 
Sunyaev-Zeldovich models, comparing the results to those obtained with texture.
Taking realistic priors for the parameters, the texture interpretation is favored, while the void and 
Sunyaev-Zeldovich hypotheses are discarded. The temperature decrement produced by voids or clusters is negligible
considering realistic values for the parameters.
\end{abstract}

\begin{keywords}
methods: data analysis - cosmic microwave background
\end{keywords}

\section{Introduction}

The study of the temperature anisotropies in the Cosmic Microwave Background (CMB) can provide crucial information on the origin
of the Universe and help us to discriminate between different cosmological models. 
Recent observations in many cosmological fields tend to support the standard inflationary model, which predicts the
anisotropies to represent an isotropic Gaussian random field. 
The Wilkinson Microwave Anisotropy Probe (WMAP) 1-year data (Bennett et al. 2003), measured the CMB fluctuations with high accuracy. 
In a first approach (Komatsu et al. 2003), these were found to be consistent with the Gaussian predictions. 
Further analyses revealed asymmetries or non-Gaussian features which have been confirmed in the 3-year WMAP data (Hinshaw et al. 2007). 
Some of these anomalies are: low multipole alignments (de Oliveira-Costa et al. 2006, Copi et al. 2007, Land \& Magueijo 2007, Chiang et al. 2007), 
north-south asymmetries (Eriksen et al. 2007),  structure alignment (Vielva et al. 2007), non-Gaussian features found with steerable (Wiaux et al. 2008) 
and directional wavelets (McEwen et al. 2006), and a very cold spot in the southern hemisphere (Cruz et al. 2007a). 
This cold spot (CS) is found at Galactic coordinates ($b = -57^\circ, l = 209^\circ$), and has an angular radius in 
the sky of $\approx 5^\circ$. 
The Spherical Mexican Hat Wavelet (SMHW) was used to detect this structure. This spherical wavelet is an optimal tool to enhance 
non-Gaussian features on the sphere. 
The first SMHW analysis (Vielva et al. 2004) was performed on 15 different wavelet scales and two estimators, skewness and kurtosis, were used.
The kurtosis deviated significantly from the range of values expected from Gaussian simulations at scales around $5^\circ$.
The spot was found to be the main cause of the deviation from Gaussianity, using the area estimator (Cruz et al. 2005).
Cruz et al. (2006) showed the incompatibility of the frequency independent signal of the CS with Galactic foregrounds, and Cay\'on, Jin \& 
Treaster (2005) introduced two new 
estimators, namely $Max$ and Higher Criticism, redetecting the CS. Cruz et al. (2007a) confirmed the detection in the 3-year WMAP data, using 
all the mentioned estimators.
In the latter paper, the probability of finding an at least as high deviation in Gaussian simulations was calculated without any
\emph{a posteriori} assumption. A conservative estimate of this probability was 1.85\%, and hence this spot is a very special feature whose origin deserves further investigation.
R{\"a}th et al.(2007) detect the CS in the 3-year WMAP data using scalar indices instead of wavelets.

Several models were proposed in order to explain this anomaly: voids (Inoue \& Silk 2006, 2007, Rudnick, Brown \& Williams 2007), 
second order gravitational effects (Tomita 2005, Tomita \& Inoue 2007), anisotropic Bianchi VII$_h$ model (Jaffe et al. 2005), finite cosmology model, (Adler, Bjorken \& Overduin 2006), asymptotically flat Lemaitre-Tolman-Bondi model (Garcia-Bellido \& Haugbolle 2008) and brane-world model (Cembranos et al. 2008) are some of them. 
Also a large-scale, non-Gaussian angular modulation (Naselsky et al. 2007) has been suggested although this hypothesis is not based on any physical model.
The Bianchi VII$_h$ model was proven by Jaffe et al. (2006) to be ruled out at high 
significance, whereas there is still no further evidence for the validity of the other explanations. A full Bayesian evidence
analysis of that Bianchi model can be found in Bridges et al. (2007a). Bridges et al. (2007b) show that the cold spot was likely to be driving any Bianchi VII$_h$ template detection.

Recently, Cruz et al. (2007b) showed that the CS could be caused by a cosmic texture. 
The amplitude and scale of the spot were consistent with that interpretation and the kurtosis of the data was compatible with
the Gaussian CMB plus texture model at all scales. The predicted number of spots at a scale of $5^\circ$ or larger, is $\approx 1$ which is consistent with the
observed spot.

In this paper we study two other possibilities, namely the Rees-Sciama (Rees \& Sciama 1968) effect caused by voids as proposed by Inoue \& Silk (2006, 2007), 
Rudnick, Brown \& Williams (2007) and the Sunyaev-Zeldovich effect produced by clusters (hereafter SZ effect, Sunyaev \& Zeldovich 1970) already 
considered in Cruz et al. (2005).
Both of them could in principle produce temperature decrements as the CS. In addition a dip (Rudnick, Brown \& Williams 2007) in the NRAO VLA Sky Survey data (NVSS, Condon et al. 1998) and the Eridanus group of galaxies (Brough et al. 2006) match approximately the size and position in the sky of the CS.
However, the values required by the parameters of either model are not very realistic. 
A Bayesian framework is very convenient to decide which model best describes the data. The prior theoretical and observational knowledge on the 
parameters is included in the Bayesian analysis and can be decisive in some cases. 
Our analysis is performed on the recently released 5-year WMAP data (Hinshaw et al. 2008). The large-scale structure of the CMB anisotropies remains almost
unchanged respect to the 3-year release (Komatsu et al. 2008).

We will first review the data and the template fitting method in section 2, then we recall in section 3 the results obtained 
for the texture model in order to compare with the SZ effect in section 4 and the void model in section 5. Finally we present our conclusions
in section 6.

\section{Bayesian template fit}

We use the combined, foreground-cleaned Q-V-W map (hereafter WCM) of the 5-year WMAP data release (Hinshaw et al. 2008).
The WCM is in the HEALPix pixelization scheme (Gorski et al. 2005) with resolution
parameter Nside = 64. Since the scale of the spot is around $5^\circ$ this resolution is sufficient
and reduces the number of pixels used in the template fitting.

The template fitting is performed in a circular area of $20^\circ$ radius centered at Galactic coordinates
($b = -57^\circ, l = 209^\circ$). 
Although the angular radius of the CS is about $5^\circ$ we have to consider at least a $20^\circ$ radius
patch to take into account the whole neighborhood of the spot for the fit. 
The SMHW convolves all the pixels in this region and they contribute in an important way to 
the detected structure.

In our template fitting we apply Bayesian statistics to the data, $\bmath{D}$, in order to find the best Hypothesis, $H$, describing the
data through a set of parameters, $\bmath{\Theta}$.

Bayes' theorem states:
\begin{equation}
P(\bmath{\Theta}|\bmath{D}, H) = \frac{P(\bmath{D}|\bmath{\Theta}, H) P (\bmath{\Theta}| H)}{P(\bmath{D}|H)},
\label{Bayes}
\end{equation} 
where $P(\bmath{\Theta}|\bmath{D},H)$ is the posterior distribution, $P(\bmath{D}|\bmath{\Theta},H)$ the likelihood, 
$P(\bmath{\Theta}|H)$ the prior and $P(\bmath{D}|H)$ the Bayesian evidence.

The null hypothesis, $H_0$, describes the data as an isotropic Gaussian random field 
(CMB) plus the noise of the WMAP data.
The alternative hypothesis, $H_1$ describes the data as CMB plus noise and an additional template, $T$, given by a physical model as for 
instance a texture spot, the SZ effect produced by a cluster or the Rees-Sciama effect generated by a local void.
$H_i$ (with $i=0,1$) depends on the parameter set $\bmath{\Theta_i}$.

For $H_i$, the probability density associated with the observed data is
\begin{equation}
P(\bmath{D}|H_i) = \int{ P(\bmath{D}|\bmath{\Theta_i},H_i)P(\bmath{\Theta_i}|H_i)d\bmath{\Theta_i} }.
\end{equation}
The evidence is hence the average of the likelihood with respect to the prior and has been extensively used for
model selection in cosmology (e.g. Jaffe 1996, Hobson, Bridle \& Lahav 2002, Hobson \& McLachlan 2003, Mukherjee, 
Parkinson \& Liddle 2005, Parkinson, Mukherjee \& Liddle 2006).
A model having a large evidence shows a high likelihood for its allowed parameter space given the data.

We redefine our notation, with the likelihood $L$, and the prior $\Pi$ so the previous integral 
for the evidence, $E$, reads:
\begin{equation}
E_{i} = \int{ L(\bmath{\Theta_i}|H_i)\Pi(\bmath{\Theta_i})d\bmath{\Theta_i} }.
\end{equation}

The likelihood function is $L \propto \exp\left(-\chi^{2}/2\right)$, where $\chi^2$ is defined as:
\begin{equation}
\chi^2 = (\bmath{D}-\bmath{T})^T \bmath{N}^{-1} (\bmath{D} - \bmath{T}),
\end{equation}
and $\bmath{N}$ is the generalised noise matrix including CMB and noise. The template $T$ depends on the parameter set $\bmath{\Theta}$.
The calculation of the noise contribution to $\bmath{N}$ is straightforward since the noise in the WCM is uncorrelated and well known.
In order to obtain the CMB part, we calculate the correlation function for the WCM taking into account the pixel and beam effects.
Comparing the resulting evidence ratios obtained from the analytical expression for the WCM correlation function with those obtained using 
simulations, the errors are negligible.

The posterior probability ratio:  
\begin{eqnarray}
 \rho \equiv  \frac{\Pr(H_1|\mathbf{D})}{\Pr(H_0|\mathbf{D})} =  \frac{E_1}{E_0} \frac{\Pr(H_1)}{\Pr(H_0)},
\label{rho}
\end{eqnarray}
can be used to decide beween both hypotheses. 
The alternative hypothesis is the more probable one when $\rho > 1$. If $\rho < 1$ the null hypothesis is the preferred one to describe the data.
The a priori probability ratio for the two models, $\Pr(H_1)/\Pr(H_0)$ is set to the fraction of the sky, $f_S$, covered by either texture, clusters or voids. In this manner we compensate the 
\emph{a posteriori} selection of the pixel where the template is centred, namely the center of the CS, ($b = -57^\circ, l = 209^\circ$).

\section{Texture}

Unified theories of high energy physics predict the production of topological defects (Kibble 1976, Vilenkin \& Shellard 2000) 
during a symmetry-breaking phase transition in the early universe. Depending on the order of the broken
symmetry, different types of defects arise, such as domain walls, strings, monopoles or texture.
Each of these extremely energetic events leave a characteristic pattern in the CMB.
For instance cosmic texture (Turok 1989) produce hot and cold spots in the CMB sky. 

Cruz et al. (2007b) showed recently that the CS could be due to a collapsing texture. 
A Bayesian template fit was performed on the 3-year WMAP data, using an approximated temperature profile based on the analytic 
result given in Turok \& Spergel (1990). 
The a priori probability ratio for the two models, $\Pr(H_1)/\Pr(H_0)$, was estimated in this case by the fraction 
of sky covered by texture, $f_S=0.017$.
The evidence ratio was $E_1 / E_0 \approx 150$, hence the resulting posterior probability ratio was $\rho = 2.5$ 
favoring the texture hypothesis versus the null hypothesis. 
The scale and amplitude of the spot ($5.1^\circ$ and $dT/T = 7.7 \times 10^{-5}$) were consistent with having being caused by a texture 
at redshift $z \sim 6$ and energy scale $\phi_0 \approx 8.7 \times 10^{15} GeV$. 
The number of texture spots of scale $5.1^\circ$ or above is around $1.1$ consistent with the single observed spot.

The amplitude of the texture spot was shown to be overestimated by a selection bias. Texture spots with amplitude $4 \times 10^{-5}$ were simulated 
and added to CMB Gaussian simulations. Performing a template fit on these single extreme events, the inferred amplitude was in average $7.9 \times 10^{-5}$, i.e. 
almost twice the true amplitude.
This means that the amplitude required by the CS is not in conflict with the upper limits on the amplitude calculated from the power 
spectrum (Urrestilla et al. 2007).

Moreover, Bevis et al. (2008) and Urrestilla et al. (2007) find a $2\sigma$ preference for a cosmological model including texture or strings, 
when considering only CMB data. However this preference is reduced when including Hubble Key Project (Freedman et al. 2001) and Big Bang 
nuclesosynthesis (Kirkman et al. 2003) constraints in the Bayesian power spectrum anlysis.

In order to estimate the significance of the $\rho = 2.5$ result, the same fitting procedure was performed on 10,000 Gaussian CMB simulations. 
The most prominent spot was selected for each simulation and a texture profile was fitted to it. Around $5.8\%$ of the simulations showed
$\rho > 2.5$ and the typical values where $\rho \approx 0.14$. This result was found to be almost independent on the priors 
or the exact shape of the approximated temperature profile.

Using the 5-year data release we obtain $E_1 / E_0 \approx 160$ and $\rho = 2.7$.
The lower noise level allows a slightly better detection of the texture template in the new data release.
We have not repeated the frequentist estimation of the significance because the result was very similar to the 3-year one.
The probability of having an at least as extreme value (i.e. $p$-value) of $\rho$ assuming that the null hypothesis is true, should be again around $5\%$.

The presence of texture would explain the excess of kurtosis observed at scales around $5^\circ$. The $1^\circ - 2^\circ$ texture predicted 
by the model would be confused with Gaussian CMB spots since the latter have their maximum power at these scales.

A number of follow-up tests can be performed to check the texture hypothesis. For instance
an all-sky template fitting looking for several texture can confirm or refute the alternative hypothesis. Also a polarization analysis
can help to decide, since Gaussian CMB spots have a correlated polarization pattern, whereas late texture do not show associated polarization.
In addition the B-mode texture power spectrum is expected to be larger than the typical inflationary one, having also a different shape (Urrestilla et al. 2007).

\section{Clusters}

The hot gas trapped in clusters modifies the CMB spectrum through the Sunyaev-Zeldovich effect. 
The high-energetic electrons present in the gas collide with the CMB photons, changing their original frequency.
Two different components can be identified. The former is produced by random thermal motion of the electrons (thermal effect),
and the second one, by the peculiar motion of the cluster with respect to the CMB (kinetic effect). 
The thermal SZ effect is typically one order of magnitude higher than the kinetic one, and therefore we focus on the first one.
The effect on the CMB photons is frequency dependent and at frequencies below 217 GHz it produces a temperature decrement.
Hence this effect is a possible candidate to explain an anomalously cold spot at the WMAP frequencies which range from 23 GHz to 94 GHz.

The temperature decrement is given by:
\begin{equation}
{\Delta T\over T_0} = g(x)y_c, 
\label{dtt}
\end{equation}
where $x=h\nu/k_B T_0$ is the frequency in dimensionless units, with $h$ and $k_B$ Planck and Boltzmann constants respectively,
and $g(x) = (x{\rm coth}(x/2) - 4)$.  
The cluster Comptonization parameter is defined as
\begin{equation}
y_c = {\sigma_T k_B \over m_e c^2} \int T_e n_e dl, 
\label{yc}
\end{equation}
where $n_e$ and $T_e$ are the intracluster electron density and temperature,
$\sigma_T$ is the Thomson cross section, and $m_e$ is the
electron mass. The integral is performed along the line of sight through the cluster. In hot clusters ($T_e > 5$ keV), the relativistic electrons lead to a slightly modified spectral behaviour of the thermal SZ effect, but this is negligible in our analysis. Also neglected is the kinetic SZ effect of the cluster.

Cruz et al. (2006) showed that the flat frequency dependence of the CS is incompatible with being caused by a SZ signal alone.
However a combination of CMB plus SZ effect may explain the spot and could have a sufficiently flat frequency spectrum.

\begin{figure}
\includegraphics[width=84mm, height=150mm]{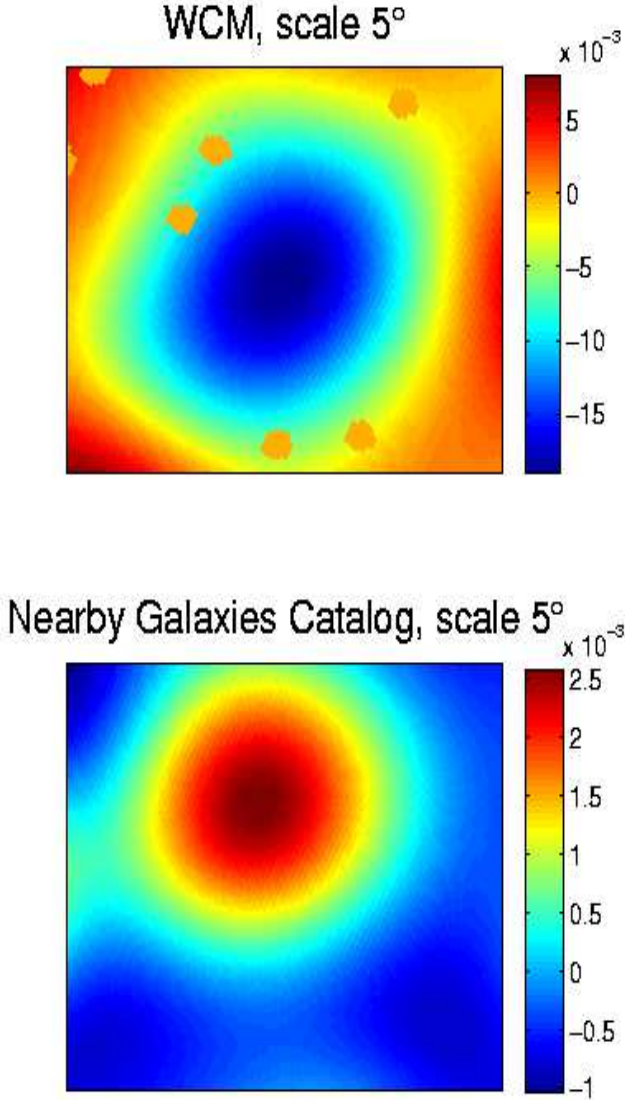}
\caption{Azimuthal projection of an $\approx 18^\circ \times 18^\circ$ patch of the WMAP combined map (WCM, top) and the Nearby Galaxies Catalog (bottom), 
centered at ($b = -57^\circ, l = 209^\circ$) and convolved with the SMHW at scale $5^\circ$. The units are wavelet coefficients of 
temperature in mK and number of counts for the WCM and Nearby Galaxies Catalog respectively.}
\label{fig:Patch}
\end{figure}

Cruz et al. (2005) already reported that a group of local galaxies were present in the direction of the CS.  They argued that the amount of
gas needed to explain the whole signal was too large for a small group of galaxies.
However the group of galaxies in the direction of the spot has been shown recently to be larger than previously thought. Brough et al. (2006),
define it as the Eridanus \emph{supergroup} with mass $M \approx 10^{14} M_\odot$. 
The angular size and position in the sky of Eridanus are similar to those of the CS, as can be seen in 
Figure~(\ref{fig:Patch}). The top panel shows the WCM data and the bottom one, the Nearby Galaxies Catalog (Tully, 1987).
Both are $\approx 18^\circ \times 18^\circ$ patches centered at ($b = -57^\circ, l = 209^\circ$) and were convolved with the SMHW at 
scale $5^\circ$. 
However there are other, more prominent nearby groups which do not have an associated CMB spot, hence this could be a statistical coincidence, 
but it is worth testing through a template fit whether there is a hidden SZ-like template in the WCM data. 

The temperature profile as a function of the angle in the sky, $\theta$, can be calculated assuming a $\beta$-model (with $\beta = 3/2$)
for the gas-density profile (see e.g. Atrio-Barandela \& M{\"u}cket, 1999):
\begin{equation}
n_e(r) = n_c(1+(r/r_c)^2)^{-3\beta/2},
\label{nr}
\end{equation}
where $n_c$ is the central electron density and $r_c$ is the core radius
of the cluster. 
Expressing the virial radius of the cluster $r_v$ in terms of the core radius as $r_v=pr_c$,
Eq.~[\ref{yc}] can be rewritten as:
\begin{equation}
y_c	= y_o\phi(\theta),
\label{ycf}
\end{equation}
where $y_o = [k_B\sigma_T/m_ec^2]r_c T n_c$, and
\begin{equation}
\phi(\theta) = {2\over\sqrt{1+(\theta/\theta_c)^2}}
\tan^{-1}\left[\sqrt{{p^2 - (\theta/\theta_c)^2}
	\over{1+(\theta/\theta_c)^2}}\right] .
\label{phi}
\end{equation}
The parameter $\theta_c$ is the angle subtended by the core radius of the cluster, whereas $p$ is not a parameter but 
a function of mass $M$ and redshift $z$. Typical values for clusters yield $p = 10 - 15$.

Summarising, our temperature profile with parameters $\theta_c$ and $y_o$ reads:
\begin{equation}
{\Delta T\over T} = g(x)y_o {2\over\sqrt{1+(\theta/\theta_c)^2}}
\tan^{-1}\left[\sqrt{{p^2 - (\theta/\theta_c)^2}
	\over{1+(\theta/\theta_c)^2}}\right]. 
\label{dttfinal}
\end{equation}
Eq.~[\ref{dttfinal}] gives the effect produced by a single 
cluster through a particular line of sight.
As the WCM is a combination of maps at different frequencies, we choose the central frequency, $\nu = 61$ GHz.

We consider two alternative hypotheses in this context. In the first one the spot would be caused by a rich cluster and in the second one by
a nearby group of the Nearby Galaxies Catalog.

We simulate the entire population of clusters with flux above 10 mJy (at 353 GHz), 
using a Monte Carlo algorithm where masses and redshifts were assigned to each cluster 
according to the Press-Schechter prescription (Press \& Schechter 1974).
We assume $\Omega_m=0.3$, $\Omega_{\Lambda}=0.7$ and $\sigma_8=0.8$ (where $\Omega_m$ and $\Omega_{\Lambda}$ are the matter and vacuum 
energy density parameters, and $\sigma_8$ the dispersion of the density field at scales of 8 Mpc). 
The cut in flux removes nearby small groups of galaxies and distant small clusters. 
The redshifts of the simulated clusters range from 0.005 to 3.4.
For each cluster with known mass and redshift the core radius is determined from the scaling relation:
\begin{equation}
 r_c = 150 M_{15}^{1/3}(1+z)^{-1} h^{-1} $kpc$
\end{equation}
where $M_{15}$ is the mass of the cluster in $10^{15} h^{-1}$ $M_{\odot}$. It is 
important to note that the cut in flux does not have any negative effect in the 
results of this paper since the fluxes we are trying to explain are much larger than 
10 mJy.
Since the redshift and core radius are known for each cluster, we calculate the angle subtended by the core, $\theta_c$.
We find the maximum angle of the 300,000 clusters to be $\theta_c \approx 0.2^\circ$. Therefore the cluster hypothesis has to
be discarded for generating a $5^\circ$ spot.

Returning to the particular case of the Eridanus supergroup found in the Nearby Galaxies Catalog, we can consider larger values of $\theta_c$, 
due to its proximity and lower values of $y_o$ since it is not a cluster but a big group or an ensemble of groups. 
In addition we assume lower values of $p$ because the gas could be rather dispersed since the gravitational attraction is lower than in big clusters.
Now the alternative hypothesis is that a big group from the Nearby Galaxies Catalog is causing the CS.

Although the center of Eridanus does not match exactly the center of the spot, we still center the template 
at ($b = -57^\circ, l = 209^\circ$), since the gas does not need to fit exactly the positions of the galaxies. This assumption
is the most favorable for the alternative hypothesis.
If the CS was due to the SZ effect we would expect high X-ray emission in the considered region. 
We can use the ROSAT all-sky survey (Snowden et al. 1997) to estimate the prior on $y_o$. The ROSAT survey excludes some 
small regions of the sky, and one of them crosses the considered patch. However the estimated average counts in the ROSAT R6 band are around 
$80$ cts s$^{-1}$ arcmin$^{-2}$. The $\beta$-models which are compatible with this background and at the same time show an extended SZ signal of several degrees, range between models with a core radius of $0.05$ Mpc and bolometric luminosities of $10^{42}$ erg/s and models with $0.15$ Mpc and bolometric luminosities of $7 \times 10^{42}$ erg/s. These models would predict at most $y_o = 10^{-7}$. 
The nearest groups in the Nearby Galaxies Catalog are at $5$ Mpc distance and as we have shown, the upper limit for the core radius is $0.15$ Mpc. In this case we get $\theta_c=0.03$ rad, which we set as the upper limit for $\theta_c$. We set the lower limit on $\theta_c$ to $1^\circ$, since the spot has $5^\circ$ radius and the pixel size is around $1^\circ$ for the chosen resolution.
Hence we consider flat priors for parameter 
$y_o$ in the interval $[0, 10^{-7}]$ and for $\theta_c$ in $[0.0175, 0.0300]$ radians. Setting $p = 5$ we obtain
a relative evidence of $ E_1 / E_0 \approx 1 $. The fraction of the sky covered by nearby galaxies in the catalog, 
is $f_S \approx 0.05$ therefore $\rho \approx 0.05 < 1$ and the SZ hypothesis is discarded again.
The best fit parameters are the upper limits of the priors, $y_o\approx 10^{-7}$ and $\theta_c=0.03$ rad. The amplitude of the temperature decrement 
does not exceed $1\mu K$ and is hence negligible compared to the deep decrement.
Therefore the $ E_1 / E_0 \approx 1 $ result is not suggesting that the SZ hypothesis is viable for explaining the spot. 
As the template has a negligible effect on the data, $\bmath{T} \approx 0$, the likelihood is almost constant in the whole parameter range and as the priors integrate to unity we have $E_1 / E_0 \approx 1$. 
This is a well-known feature of Bayesian model selection. In any problem,
if $H_1$ has additional parameters compared to $H_0$, and the likelihood
is insensitive to these parameters, then $E_1 = E_0$. 
The results do not significantly depend on the value of $p$, the chosen priors or the resolution.

The gravitational effect due to a supercluster cannot strongly affect large angular fluctuations because
these massive structures are very concentrated in a particular region. On the contrary, voids are better candidates to generate large
angular spots.

\section{Voids}

\begin{figure}
\includegraphics[width=84mm]{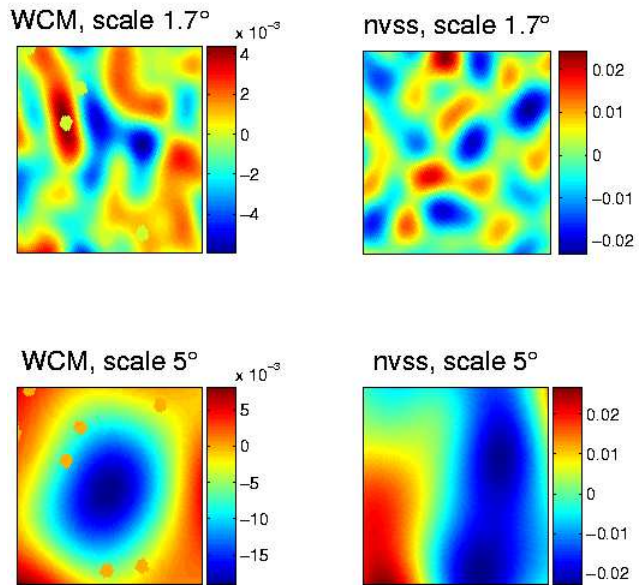}
\caption{Azimuthal projection of an $\approx 18^\circ \times 18^\circ$ patch of the WCM (left column) and the NVSS (right column), 
centered at ($b = -57^\circ, l = 209^\circ$) and convolved with the SMHW at scale $1.7^\circ$ (top row) and scale $5^\circ$ (bottom row). 
The units are wavelet coefficients of temperature in mK and radio source number counts for the WCM and NVSS respectively.}
\label{fig:WMAP_NVSS}
\end{figure}

The gravitational effect produced by cosmic voids can produce temperature decrements in the CMB and has been proposed as a possible
source of the CS (Inoue \& Silk 2006, 2007).
In analogy to the cluster case with the Eridanus supergroup, there is a dip in the NVSS radio source catalogue
that approximately matches the position of the spot. 
Rudnick, Brown \& Williams (2007), claim that the dip could be associated with a void at redshift $z < 1$ which could be the origin of  
the CS, but affirm that the significance of this coincidence is unclear.
As the fraction of the Universe filled by voids is around $30\%$ in volume at redshift $z \approx 1$ (Colberg et al. 2005), 
the fraction of the sky covered in projection, $f_S$, is close to one and hence such coincidences are not rare. 
The dip and the CS appear at different scales as can be seen in Figure \ref{fig:WMAP_NVSS} where we present a 
$\approx 18^\circ \times 18^\circ$  patch of the WMAP and NVSS data convolved with the SMHW at scales $5^\circ$ and $1.7^\circ$. 
At scale $5^\circ$ the CS reaches its maximum power in the WCM, whereas there is no significant dip visible in the NVSS data.
At scale $1.7^\circ$ where the CS is not an outstanding feature, a dip becomes visible but there are other similar dips with no
associated CMB spots, therefore the association between dip and spot is unclear and could be a coincidence. 
Moreover, there are several other dips with similar depth outside the selected region
hence the dip does not seem to be an outstanding feature of the NVSS data.
McEwen et al. (2007) found several regions where both data sets were highly correlated and one of them was located close to the CS.
This could be due to the very cold temperature of the CS which gives a high correlation with any underdense region in the NVSS data at scale $5^\circ$
and not to the relation between dip and spot since they have different scales (see Figure \ref{fig:WMAP_NVSS}).
Rudnick, Brown \& Williams (2007) also performed an order of magnitude calculation to estimate the radius of the void, $r_v$, needed to generate such a CMB spot through the 
Rees-Sciama effect obtaining $r_v \approx 200 h^{-1}$ Mpc in concordance with previous calculations (Inoue \& Silk 2007). The probability of observing such a 
void in standard theory is very small. For instance, a void with $r_v = 140 h^{-1}$ Mpc will occur with a probability of $5 \times 10^{-10}$ in an optimistic case 
(Rudnick, Brown \& Williams 2007). 
Typical observed values are $r_v \sim 10 h^{-1}$ Mpc for a matter density contrast of $\delta_m \approx -0.8$ (Hoyle \& Vogeley 2004, Patiri et al. 2006).
Similar results are found in N-body simulations (Colberg et al. 2005). 
However it is worth performing a Bayesian template fit for the void hypothesis.

Mart\'{\i}nez--Gonz\'alez et al. (1990) and Mart\'{\i}nez--Gonz\'alez \& Sanz (1990), performed a calculation of the temperature anisotropy generated by a compensated void in an Einstein-de Sitter Universe,
\begin{eqnarray}
\frac{\Delta T}{T}=\frac{16}{3}\frac{r_v^{3}}{\left(1- d\cos\alpha \right)^{3}} \sqrt{1 - \left(\frac{d}{r_v}\right)^{2} \sin^2\alpha } \times 
\nonumber
\\
\times \left(\frac{9}{2}\gamma -4 + \left(\frac{d}{r_v}\right)^{2} \sin^{2} \alpha \right)
\label{eq:dt_t_void}
\end{eqnarray}
where $\alpha$ is the angle formed between the direction of observation and the direction defined by the centre of the structure, and $d$ is the distance to the centre of the structure. They adopt the thin shell approximation, where the void is an empty spherical region surrounded by a thin shell containing the matter which would have been inside the void. Outside the void the Universe is represented by an Einstein-de Sitter model and the shell propagates approximately as $r_v \propto t^\gamma$ and $\gamma \approx 0.8$. The units are $8 \pi G = c = d_h \equiv 1$, where $d_h$ is the horizon distance. Eq.~[\ref{eq:dt_t_void}] applies to photons crossing the void. The photons that do not cross the void are unaffected since there is no net mass excess.

It is straightforward to show that Eq.~[\ref{eq:dt_t_void}] is still valid in a flat, $\Lambda$ - Cold Dark Matter ($\Lambda$-CDM) Universe, assuming that $r_v \propto t^\gamma$, and considering that
\begin{eqnarray}
d &=& \int_{0}^{z} \frac{dz}{H_0 F(z)},
\\
F(z) &=& \sqrt{\Omega_{\Lambda}+(1+z)^{3}\Omega_m+(1+z)^{4}\Omega_r},
\end{eqnarray}
where $\Omega_r = 10^{-4}$ is the radiation density parameter.
If we further assume that $r_v$ remains constant in comoving coordinates we have:
\begin{equation}
\gamma = \frac{d(\log{a})}{d(\log{t})} = F(z) \int_{z}^{\infty} \frac{d\bar{z}}{\left(1+\bar{z}\right) F(\bar{z})}
\end{equation}
with scale factor $a$.

Introducing the above expressions for $d$ and $\gamma$ in Eq.~[\ref{eq:dt_t_void}] we have our temperature template with parameters $z$, $r_v$ in a flat $\Lambda$-CDM Universe.

\begin{figure}
\includegraphics[width=84mm]{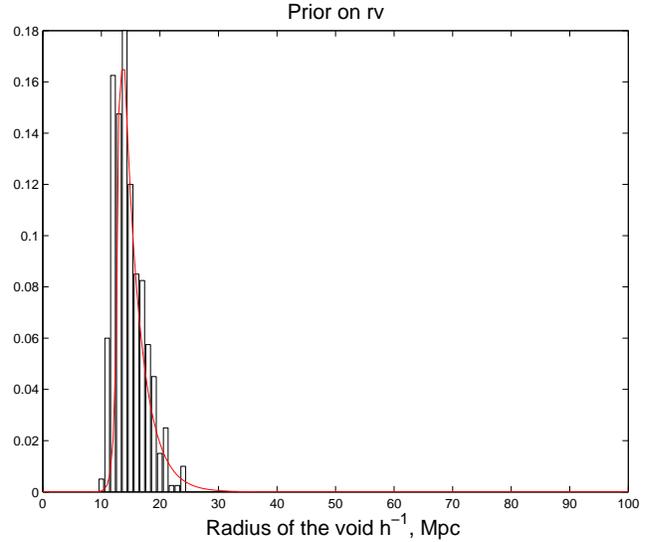}
\caption{Prior on the radius of the void parameter (red line). The underlying histogram is taken from the distribution found in the 
Two-Degree Field Galaxy Redshift Survey data (Fig. 3 in Hoyle \& Vogeley 2004).}
\label{fig:Pr_rv1}
\end{figure}

We choose the prior on $r_v$ in concordance with the void radius distribution found in the Two-Degree Field Galaxy Redshift Survey data
(Fig. 3 in Hoyle \& Vogeley 2004). We extrapolate these values for large $r_v$ through an exponential decay as shown in Figure~(\ref{fig:Pr_rv1}).
This extrapolation is generous with big voids since the derived decay is faster than the exponential used.
For the redshift, we choose a flat prior between 0 and 1. This assumption is again rather generous with large radii, since according to N-body simulations (Colberg et al. 2005) the void sizes should be smaller at high, unobserved redshifts.
We set the a priori probability ratio for the two models to unity since voids cover almost all the sky.
The obtained evidence ratio is $E_1 / E_0 \approx 1$ and therefore the posterior probability ratio is $\rho \approx 1$. 
This result is not suggesting that the void model is viable as the best fit template corresponds to $r_v = 27 h^{-1}$ Mpc and $z = 0.03$ which produces a temperature anisotropy of $0.2\mu K$.
As in the cluster case the template is completely negligible, the likelihood almost constant, and the priors integrate to unity, therefore $E_1 / E_0 \approx 1$. 
%
%
The contribution of voids with $r_v > 30 h^{-1}$Mpc to the evidence integral is negligible and the results remain unchanged varying the upper limit of 
parameter $r_v$ for values above $30 h^{-1}$Mpc. 

The big voids ($r_v > 100 h^{-1}$Mpc and $z \le 1$) described in Rudnick, Brown \& Williams (2007) which do
produce an important temperature decrement are strongly penalized by the prior, despite showing higher likelihood values. 
We redefine $H_1$ in order to test the big void hypotesis. Now only voids with $r_v > 100 h^{-1}$Mpc and $z \le 1.0$ are considered. 
We renormalise the priors in the new parameter ranges and recalculate the prior probability ratio since the big voids (in case they exist) are expected to cover only a small part of the sky.
As shown above a void with $r_v = 140 h^{-1}$ Mpc is expected to occur with a probability of $5 \times 10^{-10}$ (Rudnick, Brown \& Williams 2007). 
Hence as a valid upper limit we can assume that there is at most one big void with $r_v = 140 h^{-1}$Mpc and $z=1$.
The fraction of the sky covered by this void is $f_S \approx 0.008$.
Therefore we assume $\Pr(H_1)/\Pr(H_0) = 0.008$. Recalculating the evidence for the new parameter range, we obtain $E_1 / E_0 \approx 2$ which yields $\rho \approx 0.016$.
This result discards the possibility of explaining the CS through such a void. Hence the void hypothesis has to be rejected.

Inoue \& Silk (2006, 2007) proposed slightly different voids with very high values for the radius of the void, $r_v \sim 300 h^{-1}$ Mpc and
quasi linear density contrast values $\delta_m \approx -0.3$, while Rudnick, Brown \& Williams (2007) used $\delta_m \approx -1$.
Inoue \& Silk (2007) recognize that the required values for $r_v$ are inconsistent with observations at low $z$ and N-body simulations. 
Such a void is a $13\sigma$ result in the standard theory with $\sigma_8 = 0.74$.
They consider that a percolation process of the typically observed small voids could form the large quasi linear voids.
However these voids have not been observed yet and as already mentioned, according to N-body simulations (Colberg et al. 2005) the void sizes should be 
smaller at high, unobserved redshifts.

\section{Conclusions}

In this paper we consider different possibilities to explain a non-Gaussian cold spot found in the WMAP data. 
This spot is located at Galactic coordinates ($b = -57^\circ, l = 209^\circ$), covers around $10^\circ$ in the sky, 
has been shown to be incompatible with Gaussian simulations and
is unlikely to be produced by foregrounds (Vielva et al. 2004, Cruz et al. 2005, Cruz et al. 2006, Cruz et al. 2007a). 
We compare three alternative hypotheses to explain the CS, namely  texture, clusters and voids versus the null hypothesis. The latter 
describes the data as an isotropic Gaussian random field (CMB) plus the noise of the WMAP data. 
A Bayesian analysis takes into account the prior observational and theoretical knowledge on the parameters and can be used to
find the hypothesis best describing the data. The posterior probability ratio, $\rho$, is the quantity we use to compare the different hypotheses.
The alternative hypothesis is preferred if $\rho > 1$ and disfavored otherwise.
Cruz et al. (2007b) performed such an analysis for the texture model, finding that this model is favored versus the hypothesis
of Gaussian fluctuations with $\rho \approx 2.5$. The predicted size, amplitude, temperature profile, and number of spots generated by texture 
is consistent with the observed spot. Repeating this analysis for the 5-year WMAP data we find $\rho \approx 2.7$ due to the lower noise level.
The frequentist $p$-value associated to this result is around $5\%$.

The finding of a local supergroup (Brough et al. 2006) and a dip in the NVSS survey (Rudnick, Brown \& Williams 2007), which match the position
of the CS could be a hint for the cluster or void hypotheses, but our detailed Bayesian analysis reveals that these are statistical coincidences.
The fraction of the sky covered by clusters and voids is included in our analysis.

The large angular size and temperature decrement of the CS are hard to explain with conventional clusters or voids. 
We would need either a local, rich cluster or a void with radius $r_v \sim 100-200 h^{-1}$ Mpc located at redshift $z \le 1$.
In the cluster case only a local group with Compton parameter $y_o < 10^{-7}$ is observed in the required direction. 
The angular scale is marginally consistent with the spot but the temperature decrement generated is only a few $\mu K$ and 
hence cannot account for the CS. Our Bayesian analysis gives $\rho \approx 0.05$, discarding the cluster hypothesis.

For conventional voids with $r_v \sim 10 h^{-1}$ Mpc we obtain $\rho \approx 1$ because the likelihood is insensitive to the extra parameters. A detailed analysis reveals that these voids produce negligible temperature decrements below $1\mu K$.
Renormalising the priors for voids with radius $r_v \sim 100-200 h^{-1}$ Mpc, we obtain a posterior probability ratio of $\rho = 0.016$. 
Therefore the void hypothesis is not viable either. 
Explaining the cold spot by a void with $r_v > 100 h^{-1}$ Mpc, would introduce a $13\sigma$ anomaly in the standard model to solve the $\approx 99\%$ anomaly of the spot. The solution would be worse than the problem.

Hence we can conclude that the texture hypothesis is the most plausible explanation for the CS. 
Many scientists are uncomfortable with cosmic defects, because these failed to explain structure formation. 
However, cosmic defects are a generic consequence of symmetry breaking,
which is an essential ingredient in unified theories of particle physics. Hence it is very plausible that they exist. They might be formed at any scale below the Grand Unified scale, and would be consistent with current cosmological observations. They are even compatible with inflation. The only constraint is that they should make a subdominant contribution to structure formation (see for instance Urrestilla et al. 2007).

The virtue of the texture hypothesis is that it is fully testable. A texture would not lead to a correlated polarization pattern around the cold spot, but it would lead to a very specific gravitational lensing pattern. 
In addition there should be more CMB texture spots at scales around $1-2^\circ$. Although they are more difficult to detect because the Gaussian CMB reaches its maximum power at these scales, we may be able to detect them through an all-sky Bayesian analysis, which is already in progress.

\section*{acknowledgments}
MC, EMG and PV acknowledge financial support from the Spanish MCYT project AYA2007-68058-C03-02.
PV acknowledges financial support from the Ram\'on y Cajal project. 
We acknowledge the use of the Legacy Archive for Microwave Background Data 
Analysis (LAMBDA). Support for LAMBDA is provided by the NASA Office of Space 
Science.
This work has used the software package HEALPix (Hierarchical, Equal
Area and iso-latitude pixelization of the sphere,
http://www.eso.org/science/healpix), developed by K.M. G{\'o}rski,
E. F. Hivon, B. D. Wandelt, J. Banday, F. K. Hansen and
M. Barthelmann; the visualisation program Univiewer, developed by S.M. Mingaliev, M. Ashdown and 
V. Stolyarov; and the CAMB and CMBFAST software, developed by  A. Lewis and A. Challinor and 
by U. Seljak and M. Zaldarriaga respectively.

\end{document}
\end